\documentclass[11pt]{article}
\usepackage{amsmath}
\usepackage{cite}
\usepackage{graphicx}
\usepackage{amsfonts}
\usepackage{amssymb}
\usepackage[left=0.75in,right=0.75in,top=1in,bottom=1in]{geometry}

\providecommand{\U}[1]{\protect\rule{.1in}{.1in}}

\csname @addtoreset\endcsname{equation}{section}
\newcommand{\eq}{\begin{equation}}
\newcommand{\feq}{\end{equation}}
\newcommand{\eqn}{\begin{eqnarray}}
\newcommand{\feqn}{\end{eqnarray}}
\newcommand{\arr}{\begin{eqnarray*}}
\newcommand{\farr}{\end{eqnarray*}}

\begin{document}
\begin{titlepage}
 \begin{center}\vspace{2.0cm}
 \renewcommand{\thefootnote}{\fnsymbol{footnote}}
 {\Large{\bf Anisotropic Domain Walls and Cosmologies}}
 \vskip 1.3cm
 W. A. Sabra
 \vskip 0.2cm
  \vskip .6cm {\small{\it
  Physics Department,  American University of Beirut, Lebanon \\}}
 \end{center}
 \begin{center}
\textbf{Abstract}
\end{center} 
\vskip 0.2cm
 We investigate anisotropic domain wall and cosmological solutions in $D$-dimensional gravitational theories of arbitrary spacetime signature coupled to scalar fields with  a scalar potential. We identify a class of scalar potentials for which the second-order equations of motion admit a first-order formulation. The associated bosonic flows allow non-trivial Ricci-flat worldvolumes, while their Killing-spinor realization further requires the worldvolume to admit a non-vanishing parallel spinor.  Five-dimensional ${\cal N}=2$ gauged supergravity provides a concrete realization of this construction, with the worldvolume geometry determined by its signature and giving rise to pp-wave, hyper-K\"{a}hler, and hypersymplectic geometries. We further construct a general class of anisotropic solutions depending on a single variable, with the anisotropy encoded in a constant traceless matrix. A particular branch arises when a subset of scalar fields contribute through their kinetic terms but do not enter the scalar potential; these fields follow geodesics on the scalar manifold. As an explicit application, we recover the Kasner-dilaton-AdS solutions and show that, in the appropriate limit, they satisfy the first-order flow equations. 
  \end{titlepage}

\section{Introduction}

The study of gravitational solutions in gauged supergravity plays an
important role in high-energy physics, with applications to cosmology, the
AdS/CFT correspondence, and the Domain Wall/QFT correspondence \cite%
{Maldacena,dmqft}. Supersymmetric solutions, in particular, provide valuable
insights into the non-perturbative structure of string theory and M-theory.

In field theory and cosmology, domain walls arise as hypersurfaces
interpolating between regions characterized by different vacuum expectation
values of scalar fields \cite{vil}. They may also appear as topological
defects associated with vacua possessing non-trivial homotopy groups \cite%
{kib}. Domain wall solutions have also attracted considerable attention in
the context of brane-world scenarios \cite{braneworld} and holographic
renormalization-group flows in gauge theories \cite{r}. A comprehensive
review can be found in \cite{dworiginal}.

Most studies of supersymmetric domain walls have focused on backgrounds with
flat worldvolumes, for which the vanishing of the fermionic supersymmetry
variations leads to a system of first-order differential equations known as
flow equations. Cosmological solutions, by contrast, are commonly studied
within homogeneous and isotropic geometries. A connection between these two
settings is provided by the domain-wall/cosmology correspondence \cite{sk},
according to which a bosonic domain wall solution of a gravitational theory
with scalar potential $V$ can be continued to a cosmological solution of the
corresponding theory with scalar potential $-V$. The resulting cosmological
solutions admit pseudo-Killing spinors and can be described within the
framework of pseudo-supersymmetry.

Motivated by these developments, we investigate solutions with non-trivial
worldvolume geometries and anisotropic dependence on a single coordinate
within a common framework. We first identify a class of scalar potentials
for which the second-order equations of motion admit a first-order
formulation in arbitrary spacetime dimension and signature. The resulting
bosonic flow equations require the worldvolume of the standard domain wall
ansatz to be Ricci-flat, but not necessarily flat. When these equations
admit a Killing-spinor interpretation, the worldvolume must satisfy the
additional condition of admitting a non-vanishing parallel spinor.

We then extend the construction to anisotropic configurations depending on a
single coordinate, adapting the one-variable method originally developed for
vacuum Einstein gravity \cite{landau}. This method has also proved useful in
constructing solutions in various supergravity theories; see, for example, 
\cite{sa}. In our formulation, the anisotropic degrees of freedom are
encoded in a constant traceless matrix, while the determinant of the
worldvolume metric and the scalar fields obey a reduced system of equations.

The paper is organized as follows. We begin by developing the general
first-order formalism for gravitational theories with the proposed class of
scalar potentials in arbitrary spacetime dimension and signature. We then
specialize to five-dimensional $\mathcal{N}=2$ gauged supergravity and
investigate supersymmetric solutions with different spacetime signatures and
worldvolume geometries. The five-dimensional gauged theories can be obtained
by compactifying the exotic eleven-dimensional supergravity theories of Hull 
\cite{hull} on Calabi-Yau threefolds in the presence of internal background $%
G_{4}$-flux. Depending on the resulting five-dimensional spacetime
signature, the supersymmetry conditions lead to pp-wave, Riemannian hyper-K%
\"{a}hler, and neutral-signature hypersymplectic worldvolume geometries. We
also apply the construction to gauged supergravity theories with symmetric
scalar manifolds.

Finally, we investigate anisotropic one-variable solutions and identify
distinct classes. Of particular interest is a branch in which a subset of
scalar fields contributes through their kinetic terms but does not enter the
scalar potential. These fields follow geodesics on the scalar manifold, and
their conserved kinetic energy contributes to the Einstein equations. As an
explicit application, we recover the Kasner-dilaton-AdS solutions of \cite%
{kastor} and show that, in the appropriate limit, they satisfy the
first-order flow equations. We conclude by summarizing our results and
discussing possible directions for future research.

\section{General Solutions with Killing Spinors}

We consider $D$-dimensional gravitational theories of arbitrary spacetime
signature $(s,t)$ coupled to scalar fields. We identify a class of scalar
potentials for which the second-order field equations admit a first-order
formulation. We begin with the Lagrangian%
\begin{equation}
\mathcal{L}=\omega \left( R-2\partial _{\mu }\varphi \partial ^{\mu }\varphi
-G_{AB}\partial _{\mu }\phi ^{A}\partial ^{\mu }\phi ^{B}-2g^{2}V\right) 
\text{ },  \label{act3}
\end{equation}%
where $\omega =\sqrt{|\det g_{\mu \nu }|},$ $G_{AB}$ is the metric on the
scalar manifold, and the scalar potential is given by 
\begin{equation}
V=\frac{1}{2}e^{\beta \varphi }(D-2)^{2}\left[ G^{AB}\partial _{A}W\partial
_{B}W-\chi \frac{{\left( D-1\right) }}{(D-2)}W^{2}\right] \text{ }.
\label{flux}
\end{equation}%
Here, 
\begin{equation}
\chi =1-\frac{\beta ^{2}(D-2)}{8(D-1)}\text{ },  \label{ma}
\end{equation}%
The superpotential $W$ is a function of the scalar fields $\phi ^{A},$ and $%
\beta $ is a constant. We seek solutions with the warped metric 
\begin{equation}
ds^{2}=d\rho ^{2}+e^{2\mathcal{A}(\rho )}g_{ij}(x)dx^{i}dx^{j}\text{ },
\label{in}
\end{equation}%
and assume that all scalar fields depend only on the transverse coordinate $%
\rho .$ Throughout our work, a dot denotes differentiation with respect to $%
\rho $. The corresponding Einstein equations are%
\begin{eqnarray}
-(D-1)\left( \mathcal{\ddot{A}}+\mathcal{\dot{A}}^{2}\right) &=&2\dot{\varphi%
}^{2}+G_{AB}\dot{\phi}^{A}\dot{\phi}^{B}+{\frac{2}{D-2}}g^{2}{V}\text{ }, 
\notag \\
R_{ij} &=&e^{2\mathcal{A}}g_{ij}\left( {\frac{2}{D-2}}g^{2}{V}\text{ }%
+\left( \mathcal{\ddot{A}}+(D-1)\mathcal{\dot{A}}^{2}\right) \right) \text{ }%
,
\end{eqnarray}%
\newline
where $R_{ij}$ is the Ricci tensor of the worldvolume metric $g_{ij}$.

The scalar field equations are 
\begin{eqnarray}
{\frac{1}{\omega }}\partial _{\rho }(\omega G_{AB}\dot{\phi}^{B})-\frac{1}{2}%
\partial _{A}G_{BC}\dot{\phi}^{B}\dot{\phi}^{C}-g^{2}\partial _{A}V &=&0%
\text{ },  \notag \\
{\frac{2}{\omega }}\partial _{\rho }(\omega \dot{\varphi})-g^{2}\beta V &=&0%
\text{ }.
\end{eqnarray}%
Direct substitution shows that the complete system of second-order equations
of motion is solved by the first-order flow equations%
\begin{eqnarray}
\mathcal{\dot{A}} &=&\mp ge^{\frac{1}{2}\beta \varphi }W\text{ },\text{ \ \
\ \ \ \ }  \notag \\
\text{\ }\dot{\phi}^{A} &=&\pm (D-2)ge^{\frac{1}{2}\beta \varphi
}G^{AB}\partial _{B}W\text{ },  \notag \\
\dot{\varphi} &=&\pm \frac{\left( D-2\right) }{4}\beta ge^{\frac{1}{2}\beta
\varphi }W\text{{} },  \label{so}
\end{eqnarray}%
provided that the worldvolume metric is Ricci-flat, $R_{ij}=0.$

First-order flow equations for theories with a single scalar field and
without the additional scalar $\varphi $ were obtained in \cite{ct}. The
present construction extends this structure by including an additional
scalar $\varphi $, an arbitrary scalar-manifold metric $G_{AB}$, and
arbitrary spacetime dimension and signature. The same first-order flow
equations follow from the Killing-spinor-type system%
\begin{eqnarray}
\left( \nabla _{\mu }+\frac{1}{2}\Gamma _{\mu }ge^{\frac{1}{2}\beta \varphi
}W\right) \varepsilon &=&0\text{ },  \notag \\
\left( G_{AB}\Gamma ^{\mu }\partial _{\mu }\phi ^{B}-\left( D-2\right) ge^{%
\frac{1}{2}\beta \varphi }\partial _{A}W\right) \varepsilon &=&0\text{ }, 
\notag \\
\left( \Gamma ^{\mu }\partial _{\mu }\varphi -\frac{\beta }{4}\left(
D-2\right) ge^{\frac{1}{2}\beta \varphi }W\right) \varepsilon &=&0\text{ }.
\label{Kil2}
\end{eqnarray}%
For the domain wall ansatz, the spinor satisfies the projection condition 
\begin{equation}
\Gamma _{\rho }\varepsilon =\pm \varepsilon \text{ }.  \label{pro}
\end{equation}%
The Killing spinor then takes the form 
\begin{equation}
\varepsilon =\exp \left( \frac{\mathcal{A}(\rho )}{2}\right) \epsilon (x)%
\text{ },  \label{la}
\end{equation}%
where the worldvolume spinor satisfies 
\begin{equation}
\nabla _{i}\epsilon (x)=0\text{ }.  \label{gee}
\end{equation}%
The Killing-spinor-type formulation therefore imposes a stronger condition
on the worldvolume geometry than the bosonic first-order system.
Ricci-flatness is sufficient for the bosonic flow equations to solve the
equations of motion, whereas the Killing-spinor-type equations additionally
require a non-vanishing parallel spinor compatible with the projection
condition (\ref{pro}).

We may similarly consider cosmological solutions for which the transverse
coordinate is timelike,%
\begin{equation}
ds^{2}=-d\rho ^{2}+e^{2\mathcal{A}(\rho )}g_{ij}(x)dx^{i}dx^{j}\text{ }.
\end{equation}%
The signature and intrinsic geometry of the worldvolume metric $g_{ij}(x)$
remain unchanged; only the causal character of the transverse direction is
reversed. The corresponding solutions are obtained by replacing $g$ with $ig$
in the Killing-spinor-type equations. Since the potential enters the action
through $g^{2}V$, this continuation reverses the sign of the potential term.
The continued Killing-spinor-type equations reproduce the first-order flow
equations (\ref{so}), with the correlated projection condition 
\begin{equation}
\Gamma _{\rho }\varepsilon =\mp i\varepsilon \text{ }.
\end{equation}%
This provides the cosmological counterpart of the domain-wall construction.

\section{Supersymmetric Solutions}

We now show that five-dimensional $\mathcal{N}=2$ gauged supergravity
provides a concrete realization of the preceding construction. In this
setting, the Killing-spinor-type equations arise from the vanishing of the
fermionic supersymmetry variations in a bosonic background. We also
demonstrate how known solutions of gauged supergravity theories with scalar
fields parametrizing $SL(N,\mathbb{R})/SO(N)$ fit within the same formalism.

\subsection{Five-dimensional Gauged Supergravity}

We begin by specializing the general construction to $\mathcal{N}=2$, $D=5$
gauged supergravity coupled to $(n-1)$ Abelian vector multiplets \cite{gst2}%
. The corresponding ungauged theories with Lorentzian signature were
originally constructed in \cite{gst1}. A large class of these models arises
from compactifications of eleven-dimensional supergravity on Calabi--Yau
threefolds \cite{reduce}, while gauged versions can be obtained by
introducing internal background $G_{4}$-flux \cite{lukas,bg}. The resulting
scalar potential can also be obtained by gauging a $U(1)$ subgroup of the $R$%
-symmetry together with the axionic shift symmetry of the universal
hypermultiplet \cite{ceresole,gutsabra}.

The construction of \cite{reduce} can be extended \cite{ss} to the
eleven-dimensional supergravity theories with non-standard spacetime
signatures introduced by Hull \cite{hull}. In these theories, the four-form
kinetic term is 
\begin{equation}
\frac{\eta }{2}G_{4}\wedge \star G_{4}\text{ },
\end{equation}%
where $\eta =\pm 1,$ depending on the spacetime signature of the
eleven-dimensional theory. Upon compactification on a Calabi--Yau threefold,
this term gives rise to the five-dimensional Abelian gauge kinetic terms 
\cite{reduce}, whose overall sign is inherited from $\eta $. In the presence
of non-vanishing internal background flux, the same eleven-dimensional
kinetic term generates the scalar potential of the gauged five-dimensional
theory. Consequently, reversing the sign of the four-form kinetic term
simultaneously reverses the signs of the gauge kinetic sector and the
flux-induced scalar potential.

Alternatively, one may analytically continue the dimensionally reduced
theory to reverse the sign of the scalar potential. Such a continuation,
however, does not in general preserve a higher-dimensional interpretation in
terms of the original eleven-dimensional flux compactification.

Irrespective of the spacetime signature, the vector-multiplet scalar sector
is governed by very special real geometry \cite{vanp}. Restricting to the
vector-multiplet scalars and the hypermultiplet scalar $\varphi $ associated
with the Calabi--Yau volume, the bosonic Lagrangian takes the form (\ref%
{act3}). The vector-multiplet scalars $\phi ^{A}$ are described in terms of
constrained coordinates $X^{I}$, satisfying 
\begin{equation}
\mathcal{V}=\frac{1}{6}C_{IJK}X^{I}X^{J}X^{K}=X^{I}X_{I}=1\text{ },
\end{equation}%
where $C_{IJK}$ are real, completely symmetric constants. Useful identities
of very special geometry include 
\begin{equation}
dX_{I}=-\frac{2}{3}G_{IJ}dX^{J}\text{ },\qquad X_{I}=\frac{2}{3}G_{IJ}X^{J}%
\text{ },
\end{equation}%
and 
\begin{equation}
G^{AB}\partial _{A}X^{I}\partial _{B}X^{J}=G^{IJ}-\frac{2}{3}X^{I}X^{J}\text{
}.  \label{ide}
\end{equation}%
The flux-generated potential belongs to the family (\ref{flux}) upon
identifying 
\begin{equation}
\beta =4,\text{ \ \ \ \ }W=\alpha _{I}X^{I}\text{ },
\end{equation}%
where $\alpha _{I}$ are constant flux parameters, or equivalently gauging
parameters. Substituting these expressions into (\ref{flux}) and using the
identity (\ref{ide}), we recover the flux-generated potential \cite{bg} 
\begin{equation}
V=\frac{9}{2}e^{4\varphi }\alpha _{I}\alpha _{J}G^{IJ}\text{ }.
\end{equation}%
After consistently truncating the gauge fields while retaining the metric
and relevant scalar fields, the vanishing of the fermionic supersymmetry
variations leads to the Killing-spinor-type equations (\ref{Kil2}). The
first two equations follow from the vanishing of the gravitino and gaugino
variations, respectively, while the equation involving $\varphi $ follows
from the hyperino variation.

We may also consider the standard $U(1)$-gauged theory restricted to the
vector-multiplet sector, in which the additional scalar $\varphi $ is
absent. Within the general construction, this corresponds to removing the $%
\varphi $ sector and setting $\beta =0$. The scalar potential then belongs
to the family 
\begin{equation}
V=\frac{1}{2}(D-2)^{2}\left[ G^{AB}\partial _{A}W\partial _{B}W-\frac{{%
\left( D-1\right) }}{(D-2)}W^{2}\right] \text{ },  \label{to}
\end{equation}%
introduced in \cite{townstable}.

Depending on the theory and spacetime signature, the gauge coupling $g$ may
be real or purely imaginary, thereby realizing the two signs of the scalar
potential relevant to the solutions considered below.

\subsubsection{ $(4,1)$ signature solutions}

We first consider the theory with spacetime signature $(4,1)$. As in
previous analyses of supersymmetric domain wall solutions, we restrict
attention to Killing spinors satisfying 
\begin{equation}
\Gamma _{\rho }\varepsilon =\varepsilon \text{ }.
\end{equation}%
Rather than assuming a flat worldvolume from the outset, we allow for a
general worldvolume geometry. We employ the spinorial geometry formalism 
\cite{GranReview}, in which Dirac spinors are represented as complexified
differential forms on $\mathbb{R}^{2}$. A basis for the spinor space is
provided by $1,$ $e_{1},$ $e_{2}$ and $e_{12}=e_{1}\wedge e_{2}.$ We choose
the Killing spinor representative 
\begin{equation}
\varepsilon =\lambda 1\text{ },
\end{equation}%
together with the metric ansatz (\ref{in}), where $\lambda $ and the scalar
fields depend only on the transverse coordinate $\rho $. \ Introducing the
orthonormal frame%
\begin{equation}
\mathbf{E}^{5}=d\rho \text{ },\qquad \mathbf{E}^{i}=e^{A(\rho )}\mathbf{e}%
^{i}(x)\text{ },
\end{equation}%
the metric takes the form 
\begin{equation}
ds^{2}=(\mathbf{E}^{5})^{2}+2e^{2\mathcal{A}}(\mathbf{e}^{1}\mathbf{e}^{\bar{%
1}}+\mathbf{e}^{+}\mathbf{e}^{-})\text{ },
\end{equation}%
For real coupling $g$, the Killing spinor equations imply the flow equations 
\begin{eqnarray}
\mathcal{\dot{A}} &=&-ge^{2\varphi }W\text{ },\text{ \ \ \ \ \ \ }  \notag \\
\text{\ }\dot{X}^{I} &=&3g\alpha _{J}e^{2\varphi }\left( G^{IJ}-\frac{2}{3}%
X^{I}X^{J}\right) \text{ },  \notag \\
\dot{\varphi} &=&3ge^{2\varphi }W\text{{} }.  \label{flow}
\end{eqnarray}%
The Killing-spinor equations impose the following conditions on the
worldvolume spin connection:%
\begin{equation}
\omega _{-+}=\omega _{1\bar{1}}=\omega _{+1}=0\text{ }.
\end{equation}%
These conditions imply that the worldvolume admits a parallel null vector $%
\partial _{+}$ and is therefore locally of Brinkmann type \cite{brink,
aljose}. The spin connection conditions further restrict the worldvolume
metric to the pp-wave form 
\begin{equation}
ds_{4}^{2}=2\,dx^{+}dx^{-}+F(x^{-},x,y)(dx^{-})^{2}+dx^{2}+dy^{2}\text{ }.
\end{equation}%
These metrics admit a parallel spinor, while the bosonic Einstein equations
independently require the worldvolume to be Ricci-flat. For the metric
above, this condition reduces to%
\begin{equation}
\left( \partial _{x}^{2}+\partial _{y}^{2}\right) F=0\text{ }.
\end{equation}

For an imaginary gauge coupling, we consider solutions of the form 
\begin{equation}
ds^{2}=-d\rho ^{2}+2e^{2\mathcal{A}}(\mathbf{e}^{1}\mathbf{e}^{\bar{1}}+%
\mathbf{e}^{2}\mathbf{e}^{\bar{2}})\text{ }.
\end{equation}%
The Killing spinor equations, now subject to $\Gamma _{\rho }\varepsilon
=-i\varepsilon $, produce the same flow equations and impose the following
conditions on the Riemannian worldvolume spin connection:%
\begin{equation}
\omega _{1\bar{1}}+\omega _{2\bar{2}}=\omega _{12}=0\text{ .}
\end{equation}%
Consequently, the worldvolume is locally hyper-K\"{a}hler. Since hyper-K\"{a}
hler metrics are Ricci-flat, the worldvolume automatically satisfies the
Ricci-flatness condition required by the bosonic field equations.

\subsubsection{$(3,2)$ solutions}

We next consider supersymmetric solutions of five-dimensional supergravity
with spacetime signature $(3,2)$. We first consider solutions with a
timelike transverse coordinate and Lorentzian worldvolume. These solutions
can be obtained by analytically continuing the domain wall solutions of the $%
(4,1)$ theory. A domain wall solution of the $(4,1)$ theory with scalar
potential $V$ thus gives rise to a corresponding solution of the $(3,2)$
theory with scalar potential $-V$. This parallels the domain-wall/cosmology
correspondence, in which domain wall and cosmological solutions are related
by reversing the sign of the scalar potential. Here, however, the analytic
continuation relates theories with different spacetime signatures.

We next consider solutions of the $(3,2)$ theory with a neutral signature $%
(2,2)$ worldvolume and real gauge coupling $g$. The metric takes the form 
\begin{equation}
ds^{2}=d\rho ^{2}+2e^{2\mathcal{A}}(\mathbf{e}^{1}\mathbf{e}^{\bar{1}}-%
\mathbf{e}^{2}\mathbf{e}^{\bar{2}})\text{ }.
\end{equation}%
The Killing spinor equations, subject to $\Gamma _{\rho }\varepsilon
=\varepsilon ,$ impose the following constraints on the worldvolume spin
connection: 
\begin{equation}
\omega _{1\bar{1}}-\omega _{2\bar{2}}=\omega _{12}=0\text{ .}
\end{equation}%
Consequently, the worldvolume is locally hypersymplectic and automatically
Ricci-flat \cite{hitchin, Swann}. Hypersymplectic geometries also arise in
ungauged five-dimensional supergravity with $(3,2)$ spacetime signature, as
discussed in \cite{hsabra}.

\subsubsection{$(5,0)$ solutions}

Finally, we consider five-dimensional supergravity with Euclidean signature $%
(5,0)$. The Killing spinor equations again imply that the four-dimensional
worldvolume geometry is hyper-K\"{a}hler.

In summary, Lorentzian worldvolumes admit pp-wave geometries with
covariantly constant null Killing vectors, while Riemannian and
neutral-signature worldvolumes are hyper-K\"{a}hler and hypersymplectic,
respectively.

\subsubsection{Integrated flow equations}

We now integrate the flow equations (\ref{flow}). The first and third
equations immediately give 
\begin{equation}
e^{2\varphi }=c_{0}e^{-6\mathcal{A}}\text{ },
\end{equation}%
where $c_{0}$ is an integration constant. Using the identities of very
special geometry, the scalar flow equations can then be integrated to give%
\begin{equation}
e^{2\mathcal{A}}X_{I}=-2c_{0}g\alpha _{I}z+h_{I}\text{ },
\end{equation}%
where $h_{I}$ are integration constants and the coordinate $z$ is defined by%
\begin{equation}
\frac{dz}{d\rho }=e^{-4\mathcal{A}}\text{ }.
\end{equation}%
The corresponding metrics therefore take the form%
\begin{equation}
ds^{2}=\epsilon _{0}e^{8\mathcal{A}(z)}dz^{2}+e^{2\mathcal{A}%
(z)}g_{ij}(x)dx^{i}dx^{j}\text{ },
\end{equation}%
where the geometry of $g_{ij}(x)$ is determined by the spacetime signature
as described above, and $\epsilon _{0}=\pm 1$ specifies the causal character
of the transverse dimension.

A similar analysis applies to the $U(1)$-gauged theories. Domain wall
solutions with flat worldvolume admitting Killing spinors in these theories
were studied in \cite{cv}. The intrinsic worldvolume geometry is determined
by the spacetime signature in the same manner as above. In this case, the
scalar flow equations integrate to 
\begin{equation}
e^{2\mathcal{A}}X_{I}=-2g\alpha _{I}z+h_{I}\text{ },
\end{equation}%
where the coordinate $z$ is defined by 
\begin{equation}
\frac{dz}{d\rho }=e^{2\mathcal{A}}\text{ }.
\end{equation}%
The corresponding metric takes the form 
\begin{equation}
ds^{2}=\epsilon _{0}e^{-4\mathcal{A}(z)}dz^{2}+e^{2\mathcal{A}%
(z)}g_{ij}(x)dx^{i}dx^{j}\text{ }.
\end{equation}

\subsection{Gauged supergravity with symmetric scalar manifolds}

We next consider gauged supergravity theories in which the scalar fields
parametrize the coset space $SL(N,\mathbb{R})/SO(N)$, with Lagrangian 
\begin{equation}
\,\mathcal{L}_{D}=\omega \left( R-{{\frac{{1}}{{2}}}}\sum_{A=1}^{N-1}%
\partial _{\mu }\phi ^{A}\partial ^{\mu }\phi ^{A}-2g^{2}V\,\right) \text{ },
\end{equation}%
and scalar potential 
\begin{equation}
V=-\,\frac{1}{4}(\sum_{I=1}^{N}X_{I})^{2}+\frac{1}{2}\sum_{I=1}^{N}X_{I}^{2}%
\text{ }.  \label{pote}
\end{equation}%
The $X_{I}$ are functions of the $(N-1)$ independent scalars $\phi ^{A}$ and
satisfy 
\begin{equation}
\prod_{I=1}^{N}X_{I}=1\,.  \label{prodcon}
\end{equation}%
Further details on the structure of these theories can be found in \cite%
{cvsymmetric}. For%
\begin{equation}
N=4\frac{D-2}{D-3}\text{ },  \label{seb}
\end{equation}%
domain wall solutions were obtained in \cite{cvsymmetric} for $D=4$, $5$ and 
$7,$ corresponding respectively to $N=8$, $6,$ and $5.$ Their metrics take
the form 
\begin{equation}
ds_{D}^{2}=e^{2\mathcal{B}}\,dr^{2}\,+\,e^{2\mathcal{A}}\,\eta
_{ij}dx^{i}dx^{j}\text{ },
\end{equation}%
where 
\begin{equation}
e^{2\mathcal{A}}=(gr)^{{\frac{4}{D-3}}}\,(\prod_{I}H_{I})^{{\frac{1}{2}}-{%
\frac{2}{N}}}\text{ },\text{ \ \ }e^{2\mathcal{B}}=\frac{1}{g^{2}\,r^{2}}%
(\prod_{I}H_{I})^{-{\frac{2}{N}}}\,\,,
\end{equation}%
and the scalar fields are

\begin{equation}
X_{I}=H_{I}^{-1}\,(\prod_{J}H_{J})^{{\frac{1}{N}}}\,,\text{ \ \ \ }H_{I}=1+{%
\frac{\ell _{I}^{2}}{r^{2}}}\,\text{\ }.
\end{equation}%
To relate these solutions to our construction, we introduce $\rho $ through $%
d\rho =e^{\mathcal{B}}dr.$ Expressed in terms of the original radial
coordinate $r$, the first-order flow equations become

\begin{equation}
\frac{d\mathcal{A}}{dr}=ge^{\mathcal{B}}W\text{ },\text{ \ \ \ \ }\frac{%
d\phi ^{A}}{dr}=-2(D-2)ge^{\mathcal{B}}\delta ^{AB}\partial _{B}W\text{ },
\end{equation}%
with%
\begin{equation}
W=\frac{1}{2(D-2)}\sum_{I=1}^{N}X_{I}\ .
\end{equation}

The potential is given by (\ref{to}). The solutions of \cite{cvsymmetric}
satisfy these equations, with (\ref{seb}) precisely the condition required
for the potential (\ref{pote}) to take the form (\ref{to}). They therefore
constitute the flat-worldvolume subclass of our construction. More
generally, the flat worldvolume may be replaced by a Ricci-flat geometry.
For a Killing-spinor realization, the worldvolume must additionally admit a
non-vanishing parallel spinor.

\section{One-variable solutions}

We now construct anisotropic solutions depending on a single coordinate for
the $D$-dimensional gravitational theory described by the Lagrangian (\ref%
{act3}). We consider the metric ansatz 
\begin{equation}
ds^{2}=d\rho ^{2}+g_{ij}(\rho )dx^{i}dx^{j}\text{ }.  \label{metric}
\end{equation}%
Substituting this ansatz into the Einstein equations gives 
\begin{eqnarray}
\frac{\ddot{\omega}}{\omega }-\left( \frac{\dot{\omega}}{\omega }\right)
^{2}+\frac{1}{4}\kappa ^{i}{}_{j}\kappa ^{j}{}_{i} &=&{-}\left( 2\dot{\varphi%
}^{2}+{G}_{AB}\dot{\phi}^{A}\dot{\phi}^{B}+{\frac{2}{D-2}}g^{2}{V}\right) 
\text{ },  \label{emon} \\
\dot{\kappa}^{i}{}_{j}+\frac{\dot{\omega}}{\omega }\kappa ^{i}{}_{j} &=&-{%
\frac{4}{D-2}}g^{2}{V}\delta ^{i}{}_{j}\text{ }{,}  \label{emtw}
\end{eqnarray}%
where $\kappa _{ij}=\dot{g}_{ij}$. Taking the trace of the second equation
gives 
\begin{equation}
\frac{\ddot{\omega}}{\omega }=-2{\frac{\left( D-1\right) }{\left( D-2\right) 
}}g^{2}{V}\text{ }.  \label{detco}
\end{equation}%
Using this relation, equation (\ref{emtw}) can be integrated to give 
\begin{equation}
\kappa ^{i}{}_{j}=\theta ^{i}{}_{j}\frac{1}{\omega }+\frac{2}{D-1}\frac{\dot{%
\omega}}{\omega }\delta ^{i}{}_{j}\text{ }.  \label{som}
\end{equation}%
Here, $\theta ^{i}{}_{j}$ are the elements of a constant traceless matrix.
Consequently,

\begin{equation}
\dot{g}_{ij}=g_{ik}\theta ^{k}{}_{j}\frac{1}{\omega }+\frac{2}{D-1}\frac{%
\dot{\omega}}{\omega }g_{ij}\text{ }.  \label{metriccon}
\end{equation}%
Introducing the decomposition%
\begin{equation}
g_{ij}=\omega ^{2/(D-1)}h_{ij}\text{ },
\end{equation}%
we obtain 
\begin{equation}
\dot{h}_{ij}=\frac{1}{\omega }h_{ik}\theta ^{k}{}_{j}\text{ }.  \label{napp}
\end{equation}%
Equations (\ref{emon}), (\ref{som}), and (\ref{detco}) then imply

\begin{equation}
\frac{\left( D-2\right) }{\left( D-1\right) }\left( \frac{\ddot{\omega}}{%
\omega }-\left( \frac{\dot{\omega}}{\omega }\right) ^{2}\right) +\frac{1}{%
4\omega ^{2}}\theta ^{i}{}_{j}\theta ^{j}{}_{i}={-2}\dot{\varphi}^{2}-{G}%
_{AB}\dot{\phi}^{A}\dot{\phi}^{B}\text{ },  \label{cl}
\end{equation}%
Using equation (\ref{detco}), the equation of motion for the scalar $\varphi 
$ can be written as 
\begin{equation}
\frac{d}{d\rho }\left( \omega \dot{\varphi}\right) +\frac{\beta }{4}\frac{%
\left( D-2\right) }{\left( D-1\right) }\ddot{\omega}=0\text{ }.
\end{equation}%
Integrating once gives

\begin{equation}
\dot{\varphi}=-\frac{\beta }{4}\frac{\left( D-2\right) }{\left( D-1\right) }%
\frac{\dot{\omega}}{\omega }+\frac{c}{\omega }\text{ },  \label{int}
\end{equation}%
where $c$ is an integration constant. The equations of motion for the scalar
fields $\phi ^{A}$ are%
\begin{equation}
{\frac{1}{\omega }}\partial _{\rho }(\omega G_{AB}\dot{\phi}^{B})-\frac{1}{2}%
\partial _{A}G_{BC}\dot{\phi}^{B}\dot{\phi}^{C}-g^{2}\partial _{A}V=0\text{ }%
.
\end{equation}

\subsection{Solutions with Scalar Flow}

We first consider solutions for which the scalars $\phi ^{A}$ obey the
first-order flow (\ref{so}). We do not, however, impose the corresponding
first-order equations for $\omega $ and $\varphi $. Instead, their evolution
is determined by requiring the scalar flow to satisfy the full second-order
equations of motion. For $\chi \neq 0$, substituting the first-order flow of 
$\phi ^{A}$ into their equations of motion and using the integrated $\varphi 
$ equation (\ref{int}), gives

\begin{equation}
\frac{\dot{\omega}}{\omega }=\pm (D-1)ge^{\beta \varphi /2}W-\frac{\beta c}{%
2\chi \omega }\text{ }.
\end{equation}%
Substituting this result back into (\ref{int}) yields

\begin{equation}
\dot{\varphi}=\mp \frac{\beta }{4}(D-2)ge^{\beta \varphi /2}W+\frac{c}{\chi
\omega }\text{ }.
\end{equation}%
Thus the determinant and $\varphi $ flows acquire additional terms
proportional to the integration constant $c$. The remaining Einstein
equation imposes the algebraic constraint

\begin{equation}
\theta ^{i}{}_{j}\theta ^{j}{}_{i}=-\frac{8c^{2}}{\chi }\text{ }.
\end{equation}

The case $\chi =0$ must be treated separately. For a generic non-constant
superpotential, substituting the first-order flow of $\phi ^{A}$ directly
into their equations of motion gives

\begin{equation}
\dot{\varphi}=-\frac{2}{\beta }\frac{\dot{\omega}}{\omega }\text{ }.
\end{equation}%
Comparison with equation (\ref{int}) then requires $c=0$. The remaining
equations can be integrated to give

\begin{equation}
\frac{\dot{\omega}}{\omega }=\pm (D-1)ge^{\beta \varphi /2}W+\frac{k}{\omega 
}\text{ },\qquad \theta ^{i}{}_{j}\theta ^{j}{}_{i}=0\text{ },
\end{equation}%
where $k$ is an integration constant.

For potentials of the form (\ref{to}), the additional scalar $\varphi $ is
absent and $\beta =0$. Repeating the above analysis while imposing the
first-order flow for the scalars $\phi ^{A}$, gives the determinant flow
equation

\begin{equation}
\frac{\dot{\omega}}{\omega }=\pm (D-1)gW\text{ }.
\end{equation}%
The remaining Einstein equation then imposes%
\begin{equation}
\theta ^{i}{}_{j}\theta ^{j}{}_{i}=0\text{ }.  \label{cat}
\end{equation}%
Thus, although the determinant and scalar flows coincide with those of the
corresponding first-order system, non-trivial anisotropy is still possible
provided that the constant traceless matrix satisfies (\ref{cat}).

\subsection{Geodesic Scalar Flow}

We now consider a distinct branch in which the scalar fields $\phi ^{A}$
have non-vanishing kinetic terms but do not enter the scalar potential. We
assume that\qquad 
\begin{equation*}
V=V(\varphi )\text{ },\text{ \ \ \ \ \ }\partial _{A}V=0\text{ }.
\end{equation*}%
This assumption is compatible, in particular, with a constant superpotential 
$W=W_{0}$. The fields $\phi ^{A}$ remain dynamical and contribute to the
Einstein equations through their kinetic energy. Their equations of motion
reduce to%
\begin{equation}
\frac{1}{\omega }\frac{d}{d\rho }\left( \omega G_{AB}\dot{\phi}^{B}\right) -%
\frac{1}{2}\partial _{A}G_{BC}\dot{\phi}^{B}\dot{\phi}^{C}=0\text{ }.
\end{equation}%
These equations describe geodesic motion on the scalar manifold. Introducing
an affine parameter $\tau $ through 
\begin{equation}
d\tau =\frac{d\rho }{\omega }\text{ },
\end{equation}%
they take the standard form 
\begin{equation}
\frac{d^{2}\phi ^{A}}{d\tau ^{2}}+\Gamma ^{A}{}_{BC}\frac{d\phi ^{B}}{d\tau }%
\frac{d\phi ^{C}}{d\tau }=0\text{ },
\end{equation}%
where $\Gamma ^{A}{}_{BC}$ are the Christoffel symbols associated with $%
G_{AB}.$ The norm of the geodesic velocity is conserved, giving 
\begin{equation}
G_{AB}\dot{\phi}^{A}\dot{\phi}^{B}=\frac{q}{\omega ^{2}}\text{ },
\end{equation}%
where $q$ is a constant. The equation of motion for $\varphi $ can likewise
be integrated once to give 
\begin{equation}
\dot{\varphi}=-\frac{\beta (D-2)}{4(D-1)}\frac{\dot{\omega}}{\omega }+\frac{%
\xi }{\omega }\text{ },
\end{equation}%
where $\xi $ is an integration constant. Substitution into the remaining
Einstein equation then gives the determinant constraint

\begin{equation}
\chi \,\dot{\omega}^{\,2}+\beta \xi \,\dot{\omega}-\frac{D-1}{D-2}\left(
2\xi ^{2}+q+\frac{1}{4}\theta ^{i}{}_{j}\theta ^{j}{}_{i}\right) +\frac{%
2(D-1)}{D-2}g^{2}V\,\omega ^{2}=0\text{ }.  \label{gcon}
\end{equation}

The corresponding cosmological solutions are obtained by taking the
transverse coordinate to be timelike, 
\begin{equation}
ds^{2}=-d\rho ^{2}+g_{ij}(\rho )dx^{i}dx^{j}\text{ }.  \label{cos}
\end{equation}%
The resulting equations are related to those of the domain-wall system by
the analytic continuation which reverses the sign of the scalar potential, $%
V\rightarrow -V$. This establishes the corresponding domain-wall/cosmology
continuation for the one-variable solutions.

\subsection{Examples of One-Variable Solutions}

The preceding construction encompasses several classes of solutions in
gauged supergravity. We illustrate its application with two examples. The
first concerns theories with a two-exponential superpotential, for which the
scalar-flow construction provides a framework for extending known isotropic
solutions to anisotropic configurations. The second is the
Kasner-dilaton-AdS family, which provides an explicit example of the
solutions described in the previous section and illustrates their
first-order-flow limit.

\subsubsection{Two-exponential superpotentials}

Many consistent truncations of gauged supergravity admit a superpotential of
the form \cite{hulldomain} 
\begin{equation}
W=a_{1}e^{-\mu _{1}\phi }+a_{2}e^{\mu _{2}\phi }\text{ ,}  \label{super}
\end{equation}%
where $a_{1},$ $a_{2},$ $\mu _{1},$ and $\mu _{2}$ are constants with $\mu
_{1}$, $\mu _{2}$ positive. The corresponding scalar potential is given by (%
\ref{to}). Supersymmetric solutions for this class, including the special
cases $\mu _{1}=\mu _{2}$ with the choices $a_{1}=a_{2}$, $a_{1}=-a_{2}$ and 
$a_{1}\neq a_{2}$, were studied in \cite{dilatonic, hulldomain}. The
construction of Section 4.1 provides a framework for obtaining anisotropic
one-variable extensions of these supersymmetric flows.

\subsubsection{Kasner-dilaton-AdS solutions}

As a concrete example, we show that the Kasner-dilaton-AdS solutions of \cite%
{kastor} arise within our general construction. We consider the branch with
constant $\phi ^{A}$, for which $q=0$, and make the identifications 
\begin{equation}
\phi =2\varphi \text{ },\qquad \beta =-2a\text{ },\qquad W=W_{0}\text{ }.
\end{equation}%
The general Lagrangian then reduces to%
\begin{equation}
\mathcal{L}=\omega \left( R-\frac{1}{2}\partial _{\mu }\phi \partial ^{\mu
}\phi -2\Lambda e^{-a\phi }\right) \text{ },
\end{equation}%
where 
\begin{equation}
\Lambda =-\frac{1}{2}g^{2}(D-1)(D-2)\chi W_{0}^{2}\text{ }.
\end{equation}%
Writing $D=n+1$, the Kasner-dilaton-AdS solution takes the form 
\begin{equation}
ds^{2}=\frac{l^{2}}{r^{2}}\frac{e^{a\phi }}{F(r)}\,dr^{2}+\frac{r^{2}}{l^{2}}%
\sum_{\alpha =0}^{n-1}\epsilon _{\alpha }F(r)^{\hat{p}_{\alpha }}(dx^{\alpha
})^{2}\text{ },
\end{equation}%
with 
\begin{equation}
e^{a\phi }=\left( \frac{r}{l}\right) ^{2\sigma }F(r)^{a\hat{p}_{\phi }/2}%
\text{ },\qquad F(r)=1-\left( \frac{r_{0}}{r}\right) ^{n-\sigma }\text{ }%
,\qquad \sigma =\frac{1}{2}(n-1)a^{2}\text{ }.
\end{equation}%
Introducing the transverse coordinate $\rho $ through 
\begin{equation}
d\rho =\frac{l}{r}e^{a\phi /2}F(r)^{-1/2}\,dr\text{ },
\end{equation}%
the worldvolume metric equations give 
\begin{equation}
\sum_{\alpha =0}^{n-1}\hat{p}_{\alpha }=1+\frac{a}{2}\hat{p}_{\phi }\text{ }.
\label{hj}
\end{equation}%
Substitution into the determinant constraints (\ref{gcon}), with $q=0,$
gives the quadratic constraint 
\begin{equation}
\sum_{\alpha =0}^{n-1}\hat{p}_{\alpha }^{2}+\frac{1}{2}\hat{p}_{\phi
}^{2}=\left( 1+\frac{a}{2}\hat{p}_{\phi }\right) ^{2}.
\end{equation}%
Introducing the rescaled exponents 
\begin{equation}
\hat{p}_{\alpha }=\frac{p_{\alpha }}{1-\frac{a}{2}p_{\phi }},\qquad \hat{p}%
_{\phi }=\frac{p_{\phi }}{1-\frac{a}{2}p_{\phi }}\text{ },
\end{equation}%
the constraints take the standard Kasner form

\begin{equation}
\sum_{\alpha =0}^{n-1}p_{\alpha }=1,\qquad \sum_{\alpha =0}^{n-1}p_{\alpha
}^{2}+\frac{1}{2}p_{\phi }^{2}=1\text{ }.
\end{equation}

The first-order-flow limit is obtained by taking $r_{0}\rightarrow 0$, \ so
that $F(r)\rightarrow 1$. In this limit, the solution reduces to the
dilaton-AdS vacuum 
\begin{equation}
ds^{2}=\frac{l^{2}}{r^{2}}e^{a\phi }\,dr^{2}+\frac{r^{2}}{l^{2}}\sum_{\alpha
=0}^{n-1}\epsilon _{\alpha }(dx^{\alpha })^{2}\text{ },\qquad e^{a\phi
}=\left( \frac{r}{l}\right) ^{2\sigma }\text{ },
\end{equation}%
with 
\begin{equation}
\sigma =\frac{1}{2}a^{2}(n-1)\text{ }.
\end{equation}%
Using 
\begin{equation}
d\rho =\frac{l}{r}e^{a\phi /2}\,dr\text{ },\text{ \ \ \ }\phi =2\varphi 
\text{ ,}
\end{equation}%
we obtain 
\begin{equation}
\frac{\dot{\omega}}{\omega }=\frac{n}{l}e^{-a\varphi }\text{ },\qquad \dot{%
\varphi}=\frac{a(n-1)}{2l}e^{-a\varphi }\text{ }.  \label{pa}
\end{equation}%
These expressions agree with the first-order flow equations (\ref{so}).
Indeed, identifying $\omega =e^{(D-1)\mathcal{A}}$ and setting $%
gW_{0}=l^{-1} $, we recover both equations in (\ref{pa}). In this limit, the
anisotropy matrix and the independent scalar charge vanish, 
\begin{equation}
\theta ^{i}{}_{j}=\xi =0\text{ }.
\end{equation}

\section{Summary and Outlook}

We have developed a common framework for domain wall and cosmological
solutions in gravitational theories coupled to scalar fields. We identified
a class of scalar potentials, determined by a superpotential $W$, for which
the second-order Einstein-scalar equations admit a first-order formulation
in arbitrary spacetime dimension and signature.

For warped solutions, the bosonic first-order system requires a Ricci-flat
worldvolume, while its Killing-spinor realization additionally requires a
non-vanishing parallel spinor compatible with the spinor projection. Thus,
the first-order flow itself requires only Ricci-flatness, whereas the
Killing-spinor equations impose the additional condition that the
worldvolume admit a compatible parallel spinor.

Five-dimensional $\mathcal{N}=2$ gauged supergravity provides a concrete
realization of this structure. The first-order equations follow from the
fermionic supersymmetry variations, while the geometry of the
four-dimensional worldvolume depends on its signature. Lorentzian
worldvolumes admit pp-wave geometries with a covariantly constant null
Killing vector, whereas Riemannian and neutral-signature worldvolumes are
hyper-K\"{a}hler and hypersymplectic, respectively. The latter two
geometries are automatically Ricci-flat, while in the pp-wave case
Ricci-flatness imposes an additional condition on the wave profile. Despite
these different geometric realizations, the solutions obey the same
first-order flow equations. Analytic continuation of the gauge coupling
relates theories of different spacetime signatures while reversing the sign
of the scalar potential.

We also constructed a broader class of anisotropic one-variable solutions in
which the anisotropy is encoded in a constant traceless matrix. One class
retains the first-order scalar flow while allowing the determinant and $%
\varphi $ flows to be deformed by an additional integration constant. The
remaining Einstein equation then relates this deformation algebraically to
the anisotropy. For potentials without the additional scalar $\varphi $, the
determinant and scalar flows retain their first-order form, while
non-trivial anisotropy is possible subject to a quadratic constraint on the
constant traceless matrix. A further class arises when the scalar fields $%
\phi ^{A}$ contribute through their kinetic terms but do not enter the
scalar potential. These fields follow geodesics on the scalar manifold with
conserved velocity norm. The anisotropy, geodesic scalar charge, and
independent $\varphi $ charge then enter a single determinant constraint.
The Kasner-dilaton-AdS solutions provide an explicit example of this
construction and reduce, in the appropriate limit, to the dilaton-AdS
solution satisfying the first-order flow equations. The corresponding
cosmological solutions follow by taking the transverse coordinate to be
timelike and reversing the sign of the scalar potential.

Several extensions would be of interest. Including non-vanishing Abelian
gauge fields may lead to charged anisotropic configurations and could
provide a first-order description of Melvin-like solutions in gauged
supergravity. Further directions include investigating the global properties
and stability of these solutions, as well as systematically classifying
admissible worldvolume geometries admitting parallel spinors in different
signatures and spacetime dimensions \cite{aljose, bryant, berger, wang, Gal}.

\bigskip

\textbf{Acknowledgments}. The author thanks J. M. Figueroa-O'Farrill for
useful discussions.

\end{document}